\documentclass[aps,prd,preprint,superscriptaddress]{revtex4}
\usepackage{graphicx}

\begin{document}
\begin{flushright}
  OU-HET-645 \ \\
\end{flushright}
\vspace{0mm}
%


\title{Can the Feynman-Hellmann theorem be used to separate \\
the connected- and disconnected-diagram contributions \\
to the nucleon sigma term ?}


\author{M.~Wakamatsu and H.~Tsujimoto}
\affiliation{Department of Physics, Faculty of Science,
Osaka University, \\
Toyonaka, Osaka 560-0043, JAPAN}



\begin{abstract}
In recent lattice QCD studies, the Feynman-Hellmann theorem is often used
to estimate separate contributions of the connected and disconnected
diagrams to the nucleon sigma term.
We demonstrate through a simple analysis
within an effective model of QCD why this could be dangerous
although the theorem is naturally expected to hold for
the sum of the two contributions, i.e. the net nucleon sigma term.
\end{abstract}


\maketitle


\section{Introduction}

The nucleon sigma term $\Sigma_{\pi N}$ is believed to be a quantity
of fundamental importance in that it gives a measure of the explicit
chiral symmetry breaking of QCD. In fact, it characterizes the effect
of finite quark mass on the mass of the nucleon as
$M_N = M_0 + \Sigma_{\pi N}$, with $M_0$ being the nucleon mass in
the chiral limit.
Recently, the JLQCD collaboration reported an estimate of the
nucleon sigma term with good precision based on the
the overlap fermion action, which preserves exact chiral
symmetry and flavor symmetries on the lattice \cite{JLQCD2008}.
They estimated the separate contributions of the connected and
disconnected diagrams to the nucleon sigma term by utilizing the
Feynman-Hellmann theorem derived within the framework of the
partially quenched QCD (PQQCD), where the quarks that couple to
external sources for the asymptotic hadrons, i.e. the valence
quarks, are distinguished from those that contribute to the quark
determinant, i.e. the sea quarks. They found that the
connected diagram gives a dominant contribution to the
nucleon sigma term and the disconnected-diagram contribution to
it is fairly small.
It appears to contradict our experience within the chiral quark soliton
model (CQSM), in which we found the dominance of the Dirac-sea
quarks over the valence quarks in this special observable
\cite{DPP1989}\nocite{Wakam1992}\nocite{KWW1999}\nocite{ES2003}
\nocite{Schweitzer2003}\nocite{WO2003}-\cite{OW2004}.
(From the physical ground, the valence and Dirac-sea contributions
in the CQSM is expected to correspond to the connected- and
disconnected-diagram contributions in the lattice QCD, at least
approximately.)

What is the cause of this discrepancy ? There appears to be little
reason to suspect the validity of the Feynman-Hellmann theorem,
especially because it can be proved on quite general theoretical
postulates.
At the same time, however, one should recognize the fact that the
general proof of the theorem in textbooks of quantum mechanics is
given only for the total mass or the total
Hamiltonian. (See \cite{Mayer2003}, for instance.)
If one divides the total contribution into two parts, it is highly
nontrivial whether the theorem holds for the individual pieces
separately. 
One might think that it is not a serious problem,
since the Feynman-Hellmann theorem is anyhow expected to hold for
the net nucleon sigma term and since only the sum is
a quantity of physical interest.
However, the authors of \cite{JLQCD2008} made a semi-quenched
estimate of the strange quark content of the nucleon within the
same framework of two-flavor QCD utilizing the Feynman-Hellmann theorem,
thereby being led to a remarkable
conclusion that the $s \bar{s}$ components in the nucleon is
very small in contrast with several past estimates in the
lattice QCD \cite{FKOU1995}\nocite{DLL1996}-\cite{Gusken1999}.
Whether this estimate is justified or not may depend
on whether the use of the Feynman-Hellmann theorem for separating the
connected- and disconnected-diagram contributions to the nucleon
sigma term is justified or not.

The purpose of the present paper is to show why a naive application
of the Feynman-Hellmann theorem can be dangerous
when it is used for the separation of the nucleon sigma term
into the two pieces. 
The strategy for verifying our claim is as follows. First, we
recall the fact that, within the framework of the CQSM, we can directly
calculate the separate contributions of the valence and Dirac-sea
quarks to the nucleon sigma term, thereby confirming
that the latter is dominant over the former.
Second, we show that a naive application of the Feynman-Hellmann
theorem leads to a totally different answer from the direct calculation,
although the sum of the valence and Dirac-sea contributions are exactly
the same in the two ways of calculating the nucleon sigma term.
Next, we shall show that
careful inspection of the derivation of the theorem indicates the
necessity of a correction term, which fills up the gap between the
direct calculation and the naive application of the Feynman-Hellmann
theorem.
Finally, bearing in mind our finding in the CQSM analysis, we shall
make some remarks on the corresponding analysis of the nucleon sigma
term in the lattice QCD by using the Feynman-Hellmann theorem.

\section{The nucleon sigma term in the Chiral Quark Soliton Model}

\subsection{A direct calculation}

We begin with the effective Lagrangian of the chiral quark soliton 
model (CQSM) with an explicit chiral symmetry
breaking \cite{DPP1989},\cite{Wakam1992} : 
\begin{equation}
 {\cal L} \ = \ {\cal L}_0 \ + \ {\cal L}^{\prime} ,
\end{equation}
where ${\cal L}_0$ denotes the chiral symmetric part given by
\begin{equation}
 {\cal L}_0 \ = \ \bar{\psi}(x) \,[\, i \not\!\partial \ - \  
 M \,U^{\gamma_5} (x) ] \,\psi(x) ,
\end{equation}
with $M$ being the dynamically generated quark mass, and
\begin{equation}
 U^{\gamma_5} (x) \ = \ e^{\,i \,\gamma_5 \,
 \mbox{\boldmath $\tau$} \cdot \mbox{\boldmath $\pi$} (x) / f_{\pi}} .
\end{equation}
On the other hand, 
\begin{equation}
 {\cal L}^{\prime} \ = \ - \,m_0 \,\bar{\psi}(x) \,\psi (x) ,
\end{equation}
is thought to simulate a small deviation from the chiral symmetry 
limit with $m_0$ being the bare quark mass. Note that 
the effective quark mass in the physical vacuum $(U = 1)$
is given by $\bar{M} = M + m_0$.
The model contains the parameters, $M, m_0, f_{\pi}$ and some
physical cutoffs.
Throughout the present study, we set $f_{\pi} = 93 \,\mbox{MeV}$
and $M = 375 \,\mbox{MeV}$, while the bare quark mass $m_0$ is
varied around the reference value $m_0 = 6 \,\mbox{MeV}$.
To fix the regularization parameters, we first define the effective
action $S_{eff} [U]$ through the relation
\begin{equation}
 Z \ = \ \int \,{\cal D} \pi \,{\cal D} \psi \,{\cal D} \psi^{\dagger}
 \,\, e^{\,i \,\int \,d^4 x \,{\cal L}} \ = \ 
 \int \,{\cal D} \pi \,\,e^{\,i \,S_{eff} [U]} .
\end{equation}
Next, to get rid of ultraviolet divergences contained in this
definition, we introduce the regularized effective action in the
proper-time regularization scheme by
\begin{equation}
 S_{eff}^{reg} [U] \ = \ \frac{1}{2} \,\,i \,N_c \,
 \int_0^{\infty} \,\frac{d \tau}{\tau} \,\,
 \varphi (\tau) \,\,\mbox{Sp} \left(\,
 e^{\,- \,\tau D^{\dagger} \,D} \ - \ e^{\,- \,\tau \,D_0^{\dagger}
 \,D_0} \,\right) ,
\end{equation}
with
\begin{equation}
 D \ = \ i \,\not\!\partial \ - \ M \,U^{\gamma_5} \ - \ m_0, \ \ \ \ \ 
 D_0 \ = \ i \,\partial \ - \ (M + m_0) . 
\end{equation}
The regularization function $\varphi (\tau)$ is introduced so as to 
cut off divergences appearing as a singularity at $\tau = 0$.
For determining it, we require that the regularized theory reproduce  
the correct normalization of the pion kinetic term as well as the mass 
term. Using the standard derivative-expansion technique,
this gives two conditions : 
\begin{eqnarray}
  \frac{N_c \,M^2}{4 \,\pi^2} \,\int_0^{\infty} \,\frac{d \tau}{\tau}
  \,\varphi (\tau) \,e^{\,- \,\tau \,\bar{M}^2} &=& f_{\pi}^2 , \\
  m_0 \cdot \frac{N_c \,M}{2 \,\pi^2 \,f_{\pi}^2} \,\int_0^{\infty}
  \,\frac{d \tau}{\tau^2} \,\varphi (\tau) \,
  e^{\,- \,\tau \,\bar{M}^2} &=& m_{\pi}^2 .
\end{eqnarray}
Since Schwinger's original choice 
$\varphi (\tau) = \theta (\tau - 1/ \Lambda^2)$, 
with $\Lambda$ being a physical cutoff energy, cannot satisfy the 
above two conditions simultaneously, we use a slightly
more complicated form as \cite{BDGPPP1993},\cite{KWW1999}
\begin{equation}
 \varphi (\tau) \ = \ c \,\,\theta 
 \left(\,\tau - 1 \,/ \,\Lambda_1^2 \,\right)
 \ + \ (1 - c) \,\,\theta \left(\,\tau - 
 1 \,/ \,\Lambda_2^2 \,\right)
\end{equation}
with $c = 0.720, \,\Lambda_1 = 412.79 \,\mbox{MeV}$ and 
$\Lambda_2 = 1330.60 \,\mbox{MeV}$.
The soliton construction in the CQSM starts with a static mean-field 
configuration of hedgehog shape as \cite{DPP1988},\cite{WY1991}
\begin{equation}
 U_0^{\gamma_5} (\mbox{\boldmath $x$}) \ = \ 
 e^{\,i \,\gamma_5 \,\mbox{\boldmath $\tau$} \cdot 
 \hat{\mbox{\boldmath $r$}} F(r)} .
\end{equation}
The quark field in this mean-field obeys the Dirac equation : 
\begin{equation}
 H \,| n \rangle \ = \ E_n \,| n \rangle ,
\end{equation}
with
\begin{equation}
 H \ = \ \frac{\mbox{\boldmath $\alpha$} \cdot \nabla}{i} 
 \ + \ M \,\beta \,e^{\,i \,\gamma_5 \,
 \mbox{\boldmath $\tau$} \cdot \hat{\mbox{\boldmath $r$}} \,F (r)}
 \ + \ m_0 \,\beta.
\end{equation}

\begin{figure}[htb] \centering
\begin{center}
\includegraphics[width=2.3in]{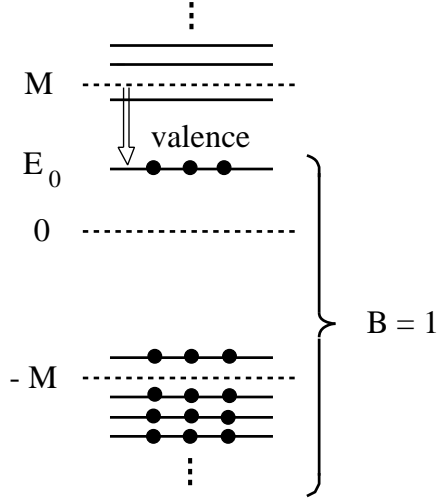}
\caption{The schematic energy spectra of the Dirac equation with
the hedgehog mean field
$U_0^{\gamma^5} (\mbox{\boldmath $x$}) = 
e^{i \,\gamma_5 \,\mbox{\boldmath $\tau$} \cdot 
\hat{\mbox{\boldmath $r$}} \,F(r)}$.}
\label{Quark_Hedgehog}
\end{center}
\end{figure}

A characteristic feature of this Dirac equation is that one deep 
(single-quark) bound state appears from the positive-energy 
Dirac continuum. We call it the valence quark orbital. An object with
 baryon number one with respect to the physical vacuum is obtained by 
putting $N_c \,(\,= \,3)$ quarks into this valence orbital as well as
all the negative-energy (Dirac-sea) orbitals. 
Accordingly, the total energy of this baryon-number-one system
is given by
\begin{equation}
 E_{static} [U] \ = \ E_{val} [U] \ + \ E_{sea} [U] .
\end{equation}
Here $E_{val}$ represents the valence quark contribution to the static energy, i.e.
\begin{equation}
 E_{val} [U] \ = \ N_c \,E_0 [U] ,
\end{equation}
with $E_0$ being the eigen-energy of the valence quark level.
On the other hand, $E_{sea}$ stands for the energy of the 
polarized Dirac sea.  
Regularizing it in the proper-time scheme, we have 
\begin{equation}
 E_{sea} [U] \ = \ \frac{N_c}{2} \,\frac{1}{\sqrt{4 \pi}} \,
 \int_0^{\infty} \,\frac{d \tau}{\tau \sqrt{\tau}} \,\,\varphi (\tau)
 \left[\,\sum_n \,e^{\,- \,\tau \,E_n^2} \ - \ 
 \sum_k \,e^{\,- \,\tau \,\epsilon_k^2} \,\right] .
\end{equation}
The energy of the physical vacuum ($U = 1$) is subtracted here with 
$\epsilon_k$ being the eigen-energy of the vacuum Hamiltonian
$H_0 \equiv H [ U \rightarrow 1]$.
The most probable pion-field configuration (or the self-consistent 
mean field) is determined by requiring the stationary condition for  
the total energy,
\begin{equation}
 \frac{\delta}{\delta F (r)} \,E_{static} [F (r) ] \ = \ 0 .
\end{equation}
This Hartree problem with infinitely many Dirac-sea orbital can be
solved by using the numerical method of Kahana, Ripka and
Soni \cite{KR1984},\cite{KRS1984}.
It also enables us to evaluate any nucleon observables with
full inclusion of the Dirac-sea quarks. 
Of our particular interest here is the nucleon sigma term, which is
defined as a $\sum_{\pi N} = m_0 \,\bar{\sigma}$ with $m_0$ being
the current quark mass and $\bar{\sigma}$ being the nucleon scalar
charge given
as $\bar{\sigma} = \langle N | \bar{u} u + \bar{d} d | N \rangle$.
By taking care of the consistency with the basic equation of motion of 
the model, the regularized expression for the nucleon scalar charge 
$\bar{\sigma}$ is given as 
\begin{equation}
 \bar{\sigma} \ = \ \bar{\sigma}_{val} \ + \ \bar{\sigma}_{sea},
\end{equation}
where
\begin{eqnarray}
 \bar{\sigma}_{val} &=& N_c \,\langle 0 | \sigma^0 | 0 \rangle , \\
 \bar{\sigma}_{sea} &=& - \,\frac{N_c}{2} \,\sum_n \,{\cal F} (E_n)
 \,\langle n | \sigma^0 | n \rangle  \ - \ (\mbox{vacuum subtraction}) ,
\end{eqnarray}
with the regularization function,
\begin{equation}
 {\cal F} (E_n) \ = \ \frac{1}{\sqrt{\pi}} \,\int_0^{\infty} \,
 \frac{d \tau}{\sqrt{\tau}} \,\,\varphi (\tau) \,E_n \,e^{-E_n^2 \tau} .
\end{equation}
Numerically, we find that
\begin{equation}
 \bar{\sigma} \ \simeq \ 6.86 ,
\end{equation}
with
\begin{equation}
 \bar{\sigma}_{val} \ \simeq \ 1.91, \ \ \  
 \bar{\sigma}_{sea} \ \simeq \ 4.95 ,
\end{equation}
which clearly show the dominance of the Dirac-sea contribution over 
the valence quark one. (With the choice $m_0 = 6 \,\mbox{MeV}$,
the above nucleon scalar charge gives
$\sum_{\pi N} \simeq 41.2 \,\mbox{MeV}$.
We recall that the 
nucleon scalar charge, especially its Dirac-sea contribution, is a
quantity which is extremely sensitive to the regularization scheme.
The Pauli-Villars regularization scheme, which is also used frequently 
in the CQSM, leads to much larger nucleon sigma term ranging from 
$48 \,\mbox{MeV}$ to $72 \,\mbox{MeV}$ depending on the bare quark
mass $m_0$ \cite{KWW1999},\cite{OW2004}.)
Anyhow, the predictions of the CQSM shown above appears to be 
qualitatively consistent with the results of the older simulations
in quenched lattice QCD by based on the Wilson quark
action \cite{FKOU1995}\nocite{DLL1996}-\cite{Gusken1999}.
(These old calculations by using the Wilson-type fermion, which
violate the chiral symmetry on the lattice, were criticized, however.
The criticism is that such calculations can give rise to a significant
lattice artifacts in the sea quark content arising from the
sea quark mass dependence of the additive mass renormalization
and lattice spacing \cite{MMH2002},\cite{JLQCD2008}.)
It however appears to contradict the recent results of
JLQCD collaborations 
by utilizing the Feynman-Hellmann theorem within the framework of the 
overlap fermion, which indicates the dominance of the contribution
of the connected diagram over that of the disconnected
one \cite{JLQCD2008}.
What is the cause of this discrepancy?
To answer this question, we think it useful
to evaluate the nucleon scalar charge by 
utilizing the Feynman-Hellmann theorem within the same CQSM.

\subsection{A naive application of the Feynman-Hellmann theorem}

We begin with the general statement of the Feynman-Hellmann theorem.
The theorem states that
\begin{equation}
 \frac{\partial}{\partial \alpha} \,E \ = \ 
 \langle \Psi (\alpha) \,| \, 
 \frac{\partial H (\alpha)}{\partial \alpha} \,| \,
 \Psi (\alpha) \rangle ,
\end{equation}
where
\begin{itemize}
\item $H(\alpha)$ is a Hamiltonian operator depending on a continuous 
parameter $\alpha$.
\item $|\, \Psi (\alpha) \rangle$ is an eigenstate of the Hamiltonian, 
depending implicitly upon $\alpha$.
\item $E$ is the eigen-energy of the Hamiltonian $H (\alpha)$.
\end{itemize}

In our present application, the bare quark mass $m_0$ plays the role 
of the parameter $\alpha$, and the Hamiltonian is given by
\begin{equation}
 H (m_0) \ = \ \frac{\mbox{\boldmath $\alpha$} \cdot \nabla}{i}
 \ + \ M \,\beta \, 
 e^{\,i \,\gamma_5 \,\mbox{\boldmath $\tau$}_i \cdot 
 \hat{\mbox{\boldmath $r$}} \,F (r)} \ + \ m_0 \,\beta .
 \label{Eq:DiracH}
\end{equation}
Thus, we obtain 
\begin{equation}
 \frac{\partial H}{\partial m_0} \ = \ \beta \ = \ \gamma^0 .
\end{equation}
On the other hand, the eigenstate are given as 
\begin{equation}
 |\, \Psi(m_0) \rangle \ = \ 
 \prod_{n \in occ} \,a_n^{\dagger} \,\,| vac \rangle ,
\end{equation}
where $a_n^{\dagger}$ represents the creation operator that creates a 
quark in the single-quark eigenstate $|n \rangle$ of the Hamiltonian 
$H (m_0)$, while $| vac \rangle$ is the corresponding empty vacuum. 
The Feynman-Hellmann theorem then dictates that 
\begin{equation}
 \frac{\partial}{\partial m_0} \,E(m_0) \ = \ 
 \langle \Psi (m_0) \,|\, \gamma^0 \,|\, \Psi (m_0) \rangle  .
\end{equation}
Since the r.h.s. is nothing but the scalar charge $\bar{\sigma}$ of 
the nucleon, we immediately get
\begin{equation}
 \bar{\sigma} \ = \ \frac{\partial}{\partial m_0} \,E (m_0) ,
\end{equation}
which is the anticipated result.
Remembering that the total energy is given as the sum of 
the energy of the valence quarks and that of the Dirac-sea quarks, 
one would further expect that
\begin{equation}
 \bar{\sigma} \ = \ \bar{\sigma}_{val} \ + \ \bar{\sigma}_{sea},
\end{equation}
with 
\begin{eqnarray}
 \bar{\sigma}_{val} &=& \frac{\partial}{\partial m_0} \,E_{val} (m_0), 
 \label{Eq:sbarFHval} \\
 \bar{\sigma}_{sea} &=& \frac{\partial}{\partial m_0} \,E_{sea} (m_0).
 \label{Eq:sbarFHsea}
\end{eqnarray}
Within the CQSM, we can solve the eigenvalue problem for any value of 
$m_0$ to obtain $E_{val}$ and $E_{sea}$ as functions of $m_0$, so that 
we can readily calculate the r.h.s. of (\ref{Eq:sbarFHval}) and
(\ref{Eq:sbarFHsea}).
(The mean field, or the soliton profile function $F(r)$, is
fixed throughout this calculation.)
In that way, we obtain
\begin{equation}
 \bar{\sigma} \ \simeq \ 6.87 ,
\end{equation}
with
\begin{equation}
 \bar{\sigma}_{val} \ \simeq \ 11.18, \ \ \ 
 \bar{\sigma}_{sea} \ = \ - \,4.31 .
\end{equation}
This should be compared with the answer of the direct calculation of
the nucleon scalar charge described in the previous subsection : 
\begin{equation}
 \bar{\sigma}^{(D)} \ = \ \bar{\sigma}^{(D)}_{val} \ + \ 
 \bar{\sigma}^{(D)}_{sea} \ \simeq \ 1.91 \ + \ 4.95 \ \simeq \ 6.86 .
\end{equation}
One finds that the two ways of calculating the nucleon scalar charge give 
totally different answers for the individual contributions of the valence 
and Dirac-sea quarks.
Nevertheless, both give practically the same answer
for the sum of the two contributions, i.e. for the net scalar charge
of the nucleon, or equivalently for the net nucleon sigma term.
Roughly speaking, the valence and Dirac-sea contributions in 
the CQSM corresponds to the connected and disconnected-diagram 
contributions in the lattice QCD. Then, what should be clarified
is the reason why the naive application of the
Feynman-Hellmann theorem does not reproduce
a correct answer for the individual contributions of the valence and Dirac-sea quarks to the nucleon
scalar charge, even though the net result, i.e. the sum of them, is correctly reproduced.

\subsection{A careful treatment and a resolution of the puzzle}

To reveal the origin of the discrepancy above, we first recall
a general proof of the Feynman-Hellmann theorem in the form
convenient for our discussion below.
We start with the expression of the energy given as
\begin{equation}
 E (\alpha) \ = \ \langle \Psi (\alpha) \,| \,
 H (\alpha) \,| \,\Psi (\alpha) \rangle .
\end{equation}
Here, we assume that $| \Psi (\alpha) \rangle$ is normalized as
$\langle \Psi (\alpha) \,| \,\Psi (\alpha) \rangle = 1$. 
For the proof of the theorem, the state $| \Psi (\alpha) \rangle$
need not be an exact
eigenstate of the Hamiltonian $H (\alpha)$.
For instance, it can be an approximate eigenstate in
Hatree-Fock theory, which is variationally optimized with respect
to the Hamiltonian \cite{Mayer2003}. Under a small variation of a
parameter $\alpha$, the change of $E (\alpha)$ is given by
\begin{eqnarray}
 \delta E (\alpha) &=& \langle \Psi (\alpha) \,| \,
 \delta H (\alpha) \,| \,\Psi (\alpha) \rangle \nonumber \\
 &+& \langle \Psi (\alpha) \,| \,
 H (\alpha) \,| \,\delta \,\Psi (\alpha) \rangle \ + \ 
 \langle \delta \Psi (\alpha) \,| \,
 H (\alpha) \,| \,\Psi (\alpha) \rangle .
\end{eqnarray}
If the state $| \Psi (\alpha) \rangle$ is variationaly
optimized with respect to the Hamiltonian, the 2nd line
of the above equation is expected to vanish, i.e.
\begin{equation}
 \langle \Psi (\alpha) \,| \,
 H (\alpha) \,| \,\delta \,\Psi (\alpha) \rangle \ + \ 
 \mbox{c.c.} \ = \ 0 ,
\end{equation}
where, c.c. means the complex conjugate of the 1st term.
We therefore obtain
\begin{equation}
 \delta E (\alpha) \ = \ \langle \Psi (\alpha) \,| \,
 \delta H (\alpha) \,| \,\Psi (\alpha) \rangle ,
\end{equation}
or equivalently
\begin{equation}
 \frac{\partial}{\partial \alpha} \,E (\alpha) \ = \ 
 \langle \Psi (\alpha) \,| \,
 \frac{\partial}{\partial \alpha} \,
 H (\alpha) \,| \,\Psi (\alpha) \rangle ,
\end{equation}
which proves the celebrated Feynman-Hellmann theorem.

What happens if the Hamiltonian consists of two terms as
\begin{equation}
 H (\alpha) \ = \ H_1 (\alpha) \ + \ H_2 (\alpha) .
\end{equation}
Here, we are imagining the decomposition of the total energy
into the contribution of the valence quarks and that of the
Dirac-sea quarks in the CQSM.
Note that, in the CQSM, this decomposition
can in fact be realized as follows : 
\begin{eqnarray}
 H_1 &=& \langle 0 \,| \,H \,|\, 0 \rangle \,\,
 a_0^\dagger \, a_0 , \\
 H_2 &=& \sum_{n \neq 0} \,\,\langle n \,| \,H \,|\, n \rangle \,\,
 a_n^\dagger \, a_n ,
\end{eqnarray}
by taking the eigenstates of the Dirac Hamiltonian (\ref{Eq:DiracH})
as a complete set. Now, the change of $E (\alpha)$ under the
variation $\alpha \rightarrow \alpha + \delta \alpha$ is
given as
\begin{eqnarray}
 \delta E (\alpha) &=& 
 \delta E_1 (\alpha) \ + \ \delta E_2 (\alpha) \nonumber \\
 &=& \langle \Psi (\alpha) \,| \,
 \delta H_1 (\alpha) \ + \ \delta H_2 (\alpha) \,| \,
 \Psi (\alpha) \rangle \nonumber \\
 &+&
 \langle \Psi (\alpha) \,| \,
 H_1 (\alpha) \ + \ H_2 (\alpha) \,| \,
 \delta \,\Psi (\alpha) \rangle \ + \ \mbox{c.c.}
\end{eqnarray}
Assuming that $| \Psi (\alpha) \rangle$ is variationally
optimized with respect to the total Hamiltonian, it still holds that
\begin{equation}
 \langle \Psi (\alpha) \,| \,
 H_1 (\alpha) \ + \ H_2 (\alpha) \,| \,
 \delta \,\Psi (\alpha) \rangle \ + \ 
 \mbox{c.c.} \ = \ 0 . \label{Eq:Variation}
\end{equation}
However, this does not necessarily mean that $H_1$ term and $H_2$
term separately vanish as
\begin{eqnarray}
 \langle \Psi (\alpha) \,| \,
 H_1 (\alpha) \,| \,
 \delta \,\Psi (\alpha) \rangle \ + \ 
 \mbox{c.c.} \ = \ 0 , \\
 \langle \Psi (\alpha) \,| \,
 H_2 (\alpha) \,| \,
 \delta \,\Psi (\alpha) \rangle \ + \ 
 \mbox{c.c.} \ = \ 0 .
\end{eqnarray}
What is meant by (\ref{Eq:Variation}) is only the identity : 
\begin{equation}
 \langle \Psi (\alpha) \,| \,
 H_1 (\alpha) \,| \,
 \delta \,\Psi (\alpha) \rangle \ + \ 
 \mbox{c.c.} \ = \ - \,\left[\,
 \langle \Psi (\alpha) \,| \,
 H_2 (\alpha) \,| \,
 \delta \,\Psi (\alpha) \rangle \ + \ 
 \mbox{c.c.} \,\right] . \label{Eq:Identity}
\end{equation}
On account of this observation, we therefore propose a
decomposition,
\begin{eqnarray}
 \langle \Psi (\alpha) \,|\,
 \frac{\partial H_1 (\alpha)}{\partial \alpha} \,|\,
 \Psi (\alpha) \rangle &=& 
 \frac{\partial}{\partial \alpha} \,E_1 (\alpha) \ - \ 
 \left[\,\langle \Psi (\alpha) \,| \,H_1 (\alpha) \,| \,
 \frac{\partial \Psi (\alpha)}{\partial \alpha} \rangle
 \ + \ \mbox{c.c.} \,\right] , \\
 \langle \Psi (\alpha) \,|\,
 \frac{\partial H_2 (\alpha)}{\partial \alpha} \,|\,
 \Psi (\alpha) \rangle &=& 
 \frac{\partial}{\partial \alpha} \,E_2 (\alpha) \ - \ 
 \left[\,\langle \Psi (\alpha) \,| \,H_2 (\alpha) \,| \,
 \frac{\partial \Psi (\alpha)}{\partial \alpha} \rangle
 \ + \ \mbox{c.c.} \,\right] .
\end{eqnarray}
Applying this result of general consideration to our case
of interest, we obtain
\begin{equation}
 \bar{\sigma} \ = \ \bar{\sigma}_{val} \ + \ 
 \bar{\sigma}_{sea} ,
\end{equation}
with
\begin{eqnarray}
 \bar{\sigma}_{val} &=& \bar{\sigma}^{(FH)}_{val} \ + \ 
 \delta \bar{\sigma}_{val} , \\
 \bar{\sigma}_{sea} &=& \bar{\sigma}^{(FH)}_{sea} \ + \ 
 \delta \bar{\sigma}_{sea} .
\end{eqnarray}
Here, the $\bar{\sigma}^{(FH)}$ terms correspond to the answer
obtained with naive application of the Feynman-Hellman theorem
as explained in the subsection A, i.e.
\begin{eqnarray}
 \bar{\sigma}^{(FH)}_{val} &=& \frac{\partial}{\partial m_0} \,
 E_{val} (m_0) , \\
 \bar{\sigma}^{(FH)}_{sea} &=& \frac{\partial}{\partial m_0} \,
 E_{sea} (m_0) .
\end{eqnarray}
On the other hand, the correction terms to this naive answer
is given by
\begin{eqnarray}
 \delta \bar{\sigma}_{val} &=& - \,
 \lim_{\Delta m_0 \rightarrow 0} \,
 \frac{\langle \Psi (m_0 + \Delta m_0) \,| \,H \,|\,
 \Psi (m_0 + \Delta m_0) \rangle^{val} \ - \ 
 \langle \Psi (m_0) \,| \,H \,|\,\Psi (m_0) \rangle^{val}}
 {\Delta m_0} , \ \ \ \\
 \delta \bar{\sigma}_{sea} &=& - \,
 \lim_{\Delta m_0 \rightarrow 0} \,
 \frac{\langle \Psi (m_0 + \Delta m_0) \,| \,H \,|\,
 \Psi (m_0 + \Delta m_0) \rangle^{sea} \ - \ 
 \langle \Psi (m_0) \,| \,H \,|\,\Psi (m_0) \rangle^{sea}}
 {\Delta m_0} , \ \ \ 
\end{eqnarray}
with the simplified notation $H = H (m_0)$.
We emphasize that the identity (\ref{Eq:Identity}) dictates that
$\delta \bar{\sigma}_{val}$
and $\delta \bar{\sigma}_{sea}$ are not independent but must
satisfy the constraint : 
\begin{equation}
 \delta \bar{\sigma}_{val} \ + \ \delta \bar{\sigma}_{sea}
 \ = \ 0 .
\end{equation}
That is, the above correction terms generally contribute to
both of the valence quark term and the Dirac-sea term, but
they are expected to cancel in the sum, i.e. in the net
contribution to the nucleon scalar charge, or equivalently
in the nucleon sigma term.

\vspace{3mm}
\newcommand{\lw}[1]{\smash{\lower2.ex\hbox{#1}}}
\begin{table}[htb]
\caption{The CQSM predictions for the nucleon scalar charge.
The calculation by using the Feynman-Hellmann theorem is
compared with the direct calculation.}
\label{Table:scala_rcharge}
\vspace{2mm}
\begin{center}
\renewcommand{\arraystretch}{1.0}
\begin{tabular}{cccc}
\hline\hline
 & \ \ \ valence \ \ \ & \ \ \ Dirac-sea \ \ \ &
 \ \ \ total \ \ \ \\
 \hline\hline
 \ \ $\bar{\sigma}^{(FH)}$ \ \ \ & \ 11.18 & - \,4.31 &  6.87 \\
 \hline
 \ \ $\delta \bar{\sigma}$ \ \ \ & - \,9.27 & \ 9.25 & - \,0.02 \\
 \hline
 \ \ $\bar{\sigma}^{(FH)} + \delta \bar{\sigma}$ \ \ \ 
 & 1.91 & 4.94 & 6.85 \\
 \hline\hline
 \ \ $\bar{\sigma}^{(D)}$ \ \ \ & 1.91 & 4.95 & 6.86 \\
 \hline\hline
\end{tabular}
\end{center}
\end{table}

Since all the quantities appearing in the above discussion
can be calculated explicitly within the CQSM, we can verify
whether our theoretical consideration is correct or not.
Shown in table I are the results of our numerical calculation
for the relevant quantities. The 2nd row of the table show the
contributions of the valence and Dirac-sea quarks to the
quantity $\bar{\sigma}^{(FH)}$ together with the sum of them,
while the 3rd row give the corresponding contributions of
correction term $\delta \bar{\sigma}$.
One sees that the valence quark
contribution to $\delta \bar{\sigma}$ is large and negative
but the Dirac-sea contribution to $\delta \bar{\sigma}$ has
just the same magnitude with opposite sign
(aside from very small numerical error).
The 3rd row represents the sum of $\bar{\sigma}^{(FH)}$ term
and $\delta \bar{\sigma}$ term, whereas the 4th row stands for
the answer of the direct calculation of the nucleon scalar
charge. One can clearly convince that, if one properly
takes account of the correction term $\delta \bar{\sigma}$
in addition to the term $\bar{\sigma}^{(FH)}$ naively
expected from the Feynman-Hellmann theorem, the answers of
the direct calculation is legitimately reproduced not
only for the net scalar charge but also for the individual
contributions of the valence and of the Dirac-sea quarks.

Now, we have confirmed that, within the framework of the
CQSM, the direct calculation and the indirect calculation by
utilizing the (slightly modified) Feynman-Hellmann theorem
give exactly the same answer for the decomposition of the
valence and Dirac-sea contributions to the nucleon sigma term.
The answer clearly shows the dominance of the contribution of
the Dirac-sea quarks over that of the valence quarks, in sharp
contrast to the corresponding answer of the lattice QCD simulation
with use of the Feynman-Hellmann theorem \cite{JLQCD2008}.
The lattice QCD version of the Feynman-Hellmann theorem is
derived based on the framework of partially quenched
QCD (PQQCD) \cite{JLQCD2008},\cite{FM1999},
where the valence quarks that coupled to external sources for
the asymptotic hadrons are distinguished from the sea quarks
that contribute to the quark determinant \cite{BG1994},\cite{SS2001}.
By treating the masses of the valence and sea quarks as independent
variables, the PQQCD version of Feynman-Hellmann theorem is
written down in the following form :
\begin{eqnarray}
 \frac{\partial M_N}{\partial m_{val}} &=& 
 \langle N \,| \,\bar{u} \,u \ + \ \bar{d} \,d \,|\,N \rangle_{conn} ,
 \label{Eq:LatticeFHval} \\
 \frac{\partial M_N}{\partial m_{sea}} &=& 
 \langle N \,| \,\bar{u} \,u \ + \ \bar{d} \,d \,|\,N \rangle_{disc} ,
 \label{Eq:LatticeFHsea} 
\end{eqnarray}
where the short-hand notation to omit the vacuum subtraction term
$- \,V \langle 0 | \bar{u} u + \bar{d} d | 0 \rangle$ is used for
the disconnected piece.
A general strategy for evaluating these terms are as follows.
One first generates statistically independent ensembles of gauge
field configurations at several different sea quarks masses.
After that, one measures the nucleon mass for various valence quark
masses on each of those gauge ensembles.
This in principle makes it possible to evaluate valence and
sea quark mass dependence of the nucleon mass.
The physical answer for the nucleon sigma term is then
obtained by calculating the derivatives (\ref{Eq:LatticeFHval})
and (\ref{Eq:LatticeFHsea}) of $M_N$ at
the unitary point $m_{sea} = m_{val}$.
(In practice, the simulation in the
chiral region is not economical, so that the results of simulations
in the larger quark mass region are extrapolated to obtain answers
corresponding to the chiral region with the help of the partially
quenched baryon chiral perturbation theory \cite{CS2002}.)

It appears that there is no question about this general prescription.
How can we reconcile the prediction of the lattice QCD with that
of the CQSM, then ? 
Naturally, an easy explanation is to claim that the decomposition
of the valence and Dirac-sea contributions to the nucleon sigma term
in the CQSM does not simply correspond to that of the connected and
disconnected contributions to the same quantity in the lattice QCD.
We cannot deny this possibility completely, because there is no
rigorous correspondence between the two theories and their
decompositions of the nucleon sigma term.
From a physical viewpoint, however, the discrepancy seems too large
(or more than quantitative) to accept this naive conclusion.
In our opinion, this discrepancy should be taken more seriously.
If there is any resolution to this problem, we conjecture that
it must be traced back to a difference between the treatment of
the valence and sea quarks in the CQSM and
that in the partially quenched QCD.
In our framework of the CQSM, we do not need to distinguish the
masses of the valence
and Dirac-sea quarks. They are treated on the equal footing from
the beginning to the end. On the other hand, there is an
apparent asymmetry in the treatment of the valence and sea quarks in
the PQQCD. (We are talking about the asymmetry in the treatment of
the valence and sea quarks when generating ensemble of gauge field
configuration. Ideally, all the field configurations of the nucleon
constituents, i.e. the gluon field, the valence and sea quarks,
should be determined according to a self-consistent dynamics of QCD.)
It may be certainly true that the masses of the valence and sea
quarks are taken equal at the end of calculation, and that the
physical answers obtained in this unitary limit is taken to be
physical \cite{SS2001}.
This would also apply to the PQQCD version of the Feynman-Hellmann theorem.
In consideration of fundamental importance of the problem,
however, we think it very important to check the validity of the
Feynman-Hellmann theorem in an explicit manner by carrying out a
direct calculation of the connected- and
disconnected-diagram contributions to the nucleon sigma term
within the same framework of the overlap fermion.
(The direct calculation means a calculation of the three point
functions with an insertion of the scalar density operators.)

Before ending this subsection, it may be useful to
recall one plausible argument, which strongly indicates that the
contribution of the valence quarks cannot be a
dominant term of the nucleon sigma term \cite{Wakam2009}.
As is well known, the recent analyses of the pion-nucleon scattering
amplitude favor fairly large nucleon sigma term ranging
from $50 \,\mbox{MeV}$ to $70 \,\mbox{MeV}$
\cite{Olsson2000}\nocite{PASW2002}-\cite{Sainio2002}.
Depending on the uncertainty of the average $u$- and $d$-quark
masses, this implies fairly large nucleon scalar charge
$\bar{\sigma}$ of the order of 10.
As we shall argue below, it is unlikely that such a large
value of $\bar{\sigma}$ can be explained by the contribution
of three valence quarks alone. To convince it, let us consider a
relativistic bound state of $N_c \,(\,= 3)$ quarks.
Assume that these quarks
are confined in some mean field or confining potential.
A typical example is the famous
MIT bag model. The ground state wave function of this popular model
is given as
\begin{equation}
 \psi_{g.s.} (\mbox{\boldmath $r$}) \ = \ 
 \left( \begin{array}{c}
 f(r) \,\chi_s \\
 i \,\mbox{\boldmath $\sigma$} \cdot \hat{\mbox{\boldmath $r$}}
 \,g(r) \,\chi_s \\
 \end{array} \right) ,
\end{equation}
where $f(r)$ and $g(r)$ are the radial wave functions of the
upper and lower components, while $\chi_s$ is an appropriate
spin wave function. The nucleon scalar charge in this model
is easily obtained as
\begin{equation}
 \bar{\sigma} \ = \ \langle N \,| \,
 \bar{u} \,u \ + \ \bar{d} \,d \,| \, N \rangle \ = \ 
 N_c \,\int_0^R \,\left[\,(f(r))^2 \ - \ (g(r))^2 \,\right] \,r^2 \,dr ,
\end{equation}
with $R$ the bag radius. Undoubtedly, the magnitude of this quantity
is smaller than $N_c$, since the radial functions satisfy the
normalization,
\begin{equation}
 \int_0^R \,\left[\,(f(r))^2 \ + \ (g(r))^2 \,\right] \,r^2 \,dr 
 \ = \ 1 .
\end{equation}
It is clear that this observation does not depend on the exact form of
the mean field or the confining potential, so that it is quite general.
As a consequence, for any model of the nucleon, which contains
$N_c$ valence quark degrees of freedom alone, we must conclude that there
exists a upper bound such that
\begin{equation}
 \bar{\sigma}_{val} \ < \ N_c .
\end{equation}
Our prediction in the CQSM, i.e. $\bar{\sigma}_{val} \simeq 1.91$,
as well as the direct calculation in the qeunched lattice QCD in
\cite{FKOU1995}, i.e. $\langle N \,| \,\bar{u} \,u + \bar{d} \,d \,
| \,N \rangle_{connected} \simeq 2.323(15)$, satisfy the above bound.
On the other hand, the recent estimate by the JLQCD collaboration
utilizing the Feynman-Hellmann theorem \cite{JLQCD2008}, i.e.
$\langle N \,| \,\bar{u} \,u + \bar{d} \,d \,
| \,N \rangle_{connected} \simeq 5.27(75) - 7.92(8)$,
lies outside this bound. Again, highly desirable is a direct calculation
of the nucleon sigma term within the framework of the overlap
fermion, without utilizing the Feynman-Hellmann theorem.

\section{Summary and conclusion}

In summary, we have investigated the nucleon sigma term or the
nucleon scalar charge within a simple effective model of QCD,
i.e. the chiral quark soliton model.
It was demonstrated that the naive application of the
Feynman-Hellmann theorem does not reproduce the correct answer
for the separate contributions of the valence and Dirac-sea
quarks to the nucleon sigma term, which can be obtained by the
direct calculation within the same model.
It was also shown that a careful inspection of the derivation
of the Feynman-Hellmann theorem indicates the necessity of a
correction term, which fills up the gap between the direct
calculation and the naive application of the Feynman-Hellmann
theorem. Anyhow, by using two completely independent methods of
calculation, we have confirmed that the contribution of the
Dirac-sea quarks dominates over that of the valence quarks
in this unique observable of the nucleon.

This observation however appears to contradict the corresponding
answer of the recent lattice QCD simulation by JLQCD collaboration
based on the action of overlap fermion.  
They estimated the separate contributions of the connected and
disconnected diagrams to the nucleon sigma term by utilizing
the lattice QCD version of the Feynman-Hellmann theorem, which is
derived within the scheme of PQQCD, and found that the connected-
diagram gives a dominant contribution to the nucleon sigma term
and the disconnected-diagram contribution is of secondary importance.
Although we do not have any convincing reasoning to suspect the
validity of the lattice QCD version of the Feynman-Hellmann theorem,
it is highly desirable to check the validity of it by a direct
calculation of the nucleon sigma term within the same framework
of overlap fermion.
This is especially so, because the separation of the nucleon
sigma term into the contributions of valence and sea quarks seems
to be a very delicate operation as our model analysis has shown,
and also because the direct confirmation of the theorem is of
fundamental importance to check whether the theoretical framework
of the PQQCD, which was invented for handling loops of sea quarks
in the lattice QCD, is working as it is expected.


\vspace{3mm}
\begin{acknowledgments}
We would like to thank Prof. T.~Onogi for useful discussion
about the Feynman-Hellmann theorem in lattice QCD.
This work is supported in part by a Grant-in-Aid for Scientific
Research for Ministry of Education, Culture, Sports, Science
and Technology, Japan (No.~C-21540268)
\end{acknowledgments}


\end{document}